\documentclass[aps,showpacs,preprint,superscriptaddress]{revtex4}
\usepackage{graphicx}
\usepackage{subfigure}
\usepackage{float}
\usepackage{amsmath}
\usepackage{amsfonts}
\usepackage{bm}
\usepackage{bbm}
\usepackage{txfonts}
\usepackage{array}
\usepackage{amssymb}

\begin{document}
\title{Electron-positron pair production in an oscillating Sauter potential}
\author{Li Wang}
\affiliation{Key Laboratory of Beam Technology of the Ministry of Education, and College of Nuclear Science and Technology, Beijing Normal University, Beijing 100875, China}
\author{Binbing Wu}
\affiliation{Key Laboratory of Beam Technology of the Ministry of Education, and College of Nuclear Science and Technology, Beijing Normal University, Beijing 100875, China}
\author{Baisong Xie}
\email[Corresponding author. Email: ]{bsxie@bnu.edu.cn}
\affiliation{Key Laboratory of Beam Technology of the Ministry of Education, and College of Nuclear Science and Technology, Beijing Normal University, Beijing 100875, China}
\affiliation{Beijing Radiation Center, Beijing 100875, China}

\date{\today}

\begin{abstract}
Electron-positron pair production in an oscillating Sauter potential is investigated in the framework of  the computational quantum field theory. It is found that for a Sauter potential well with oscillating width and depth simultaneously, the phase difference between them has a great impact on the number of created electron-positron pairs. Optimal values of the phase difference corresponding to different oscillation frequencies are obtained. The optimal phase difference has a strong nonlinear dependence on oscillating frequency. When the potential well changes slowly, our results can be explained by the instantaneous bound states. For the higher frequency case, however, multiphoton effect is enhanced and the Pauli blocking effect has a strong inhibitory effect on pair creations. These results can provide a theoretical reference for the experimental creation of the electron-positron pairs.
\end{abstract}

\maketitle
\section{Introduction}

Since positron was predicted theoretically by Dirac in 1928 and then confirmed experimentally by Anderson in 1933\cite{Dirac1928,Anderson1933}, theoretical researches are further developed including the effective Lagrangian formalism for vacuum polarization in strong background field \cite{Sauter1931,Heisenberg1936}. Also Schwinger \cite{Schwinger1951} obtained the famous exponential rate for the electron-positron pair creation that is proportional to $\exp(-\pi E_{cr}/E)$ in a constant electric field $E$, where $E_{cr}=10^{16}\rm{V/cm}$ is the Schwinger critical field strength which is far above the presently achieved laboratory electric field.

As the laser technology advances greatly, the electron-positron pair creation in the vacuum under strong fields has become a hot research topic. For example, the generation of pairs in a spatially uniform and temporally alternating electric field was studied by using the WKB approximation method, which opened up the understanding multiphoton mechanism that electrons in the negative continuum state can absorb multiple laser photons to produce pairs \cite{Brezin1970}. Afterwards, through the study of the electric field with an arbitrary time dependence, the explicit analytic expression of the probability was obtained by Marnov and Popov\cite{Marinov1977}. So far, several methods have been developed for the study of pairs production in strong filed, such as the worldline instanton technique\cite{Gies2005,Dunne2006,Ilderton2015,Schneider2016}, the Dirac-Heisenberg-Wigner formalism \cite{Li2015,Blinne2016,Kohlfurst2018,Olugh2019}, the quantum Vlasov equation solution method \cite{Kluger1991,Alkofer2001,JiangMin2013,Sitiwaldi2017}, the computational quantum field theory \cite{Krekora2005,Lv2013,Tang2013,Jiang2013,Liu2014,Gong2018} and so on.

Many works attempted to adopt different spatial and temporal pulse shaping or combined fields to reduce thresholds necessary to trigger pair creation or simply to increase pair creation yield. Effects of laser pulse shape and carrier envelope phase on pair production have been well studied and the results show that the super-Gaussian shape is superior to the Gaussian, and the pair number density is phase-dependent \cite{Abdukerim2013}. On the other hand, as the most popular potential well, the Sauter potential served for the pair production has been extensively researched \cite{Tang2013,Krekora2005,Lv2013,Gong2018,Kohlfurst2016,Jiang2014}. Other factors affecting pair production are also of considerable interest. For instances, Krekora \textit{et al.} simulated how a supercritical bound state is created in the long-time regime and described in the production rate of pairs by four distinct regimes \cite{Krekora2005}. Here we emphasize the importance of the static or simultaneous bound states of supercritical potential because pair creation in such a potential is usually explained in terms of these bound states, for details see Refs. \cite{Jiang2013,Lv2013,Liu2014,Liu2015}.

Motivated by the pulse shaping and phase effects of the supercritical fields mentioned above, therefore, in this paper we focus our study on the pair production in a Sauter potential well oscillating in both its width and depth directions. The effects of phase difference between them on pair creation is examined carefully by the computational quantum field theory. The result shows that the optimal phase difference is not fixed for different oscillation frequencies, which can provide a theoretical reference to future experimental creation of the vacuum pairs.

This paper is organized as follows. In Sec. II, by employing the computational quantum field theory, the Dirac equation is solved, and the number of created electrons is computed. In Sec.III, the potential well model is presented, in which the time evolution of the number of created electrons under different oscillation frequencies and phase differences is simulated. In Sec.IV, we summarize our work.

\section{Outline of computational quantum field theory}\label{section2}

In the quantum field theory, the Dirac equation is adopted to describe the evolution of a single particle
\begin{equation}\label{Eq Dirac}
i\partial \hat{\psi} \left(z,t\right) / \partial{t}=\left[c\alpha_z \hat{P}+\beta c^2+V\left(z,t\right)\right] \hat{\psi}\left(z,t\right),
\end{equation}
where $\alpha_z$ and $\beta$ are Dirac matrices, $c$ is the speed of light in vacuum, $V\left(z,t\right)$ is external field that varies with time $t$ in the $z$ direction.  We use the atomic units $\hbar=e=m_e=1$.
By introducing the creation and annihilation operators, the field operator $\hat{\psi}(z,t)$ can be decomposed as follows:
\begin{equation}\label{Eq Field Operator}
\begin{aligned}
\hat{\psi}(z,t)&=\sum_{p}\hat{b}_p(t)W_p(z)+\sum_{n}\hat{d}_n^{\dag}(t)W_n(z) \\
&=\sum_{p}\hat{b}_pW_p(z,t)+\sum_n\hat{d}_n^\dag W_n(z,t),
\end{aligned}
\end{equation}
where $p$ and $n$ denote the momenta of positive and negative energy states, $\sum_{p(n)}$ presents summation over all states with positive $($negative$)$ energy, $W_{p(n)}(z)=\langle z|p(n)\rangle$ is field-free energy eigenstate of positron $($electron$)$. Note that $W_{p(n)}(z,t)=\langle z|p(n)(t)\rangle$ satisfies the single-particle time-dependent Dirac equation Eq.(\ref{Eq Dirac}).
From Eq.(\ref{Eq Field Operator}), we obtain
\begin{equation}\label{Eq fermion operators}
\begin{aligned}
\hat{b}_p(t)&=\sum_{p'}\hat{b}_{p'}U_{pp'}(z,t)+\sum_{n'}\hat{d}_{n'}^\dag U_{pn'}(z,t),\\
\hat{d}_n^\dag(t)&=\sum_{p'}\hat{b}_{p'}U_{np'}(z,t)+\sum_{n'}\hat{d}_{n'}^\dag U_{nn'}(z,t),\\
\hat{b}_p^\dag(t)&=\sum_{p'}\hat{b}_{p'}^\dag U_{pp'}^*(z,t)+\sum_{n'}\hat{d}_{n'}U_{pn'}^*(z,t),\\
\hat{d}_n(t)&=\sum_{p'}\hat{b}_{p'}^\dag U_{np'}^*(z,t)+\sum_{n'}\hat{d}_{n'}U_{nn'}^*(z,t),
\end{aligned}
\end{equation}
where $U_{p(n)p'(n')}=\langle p(n)|U(t)|p'(n')\rangle$, and the time-ordered propagator $U(t)=\textrm{exp}\{{-i\int^t d\tau [c\alpha_z \hat{p}+\beta c^2+V(z,\tau)]}\}$.

In Eq.(\ref{Eq Field Operator}), the electronic portion of the field operator is defined as $\hat{\psi}_e(z,t)\equiv \sum_p \hat{b}_p(t)W_p(z)$. So we can obtain the probability density of created electrons by
\begin{equation}\label{Eq Density}
\begin{aligned}
\rho(z,t)&=\langle vac|\hat{\psi}_e^\dag (z,t)\hat{\psi}_e(z,t)|vac\rangle \\
&=\sum_n |\sum_p U_{pn}(t)W_p(r)|^2
\end{aligned}
\end{equation}
By integrating this expression over space, the number of created electrons can be obtained as
\begin{equation}\label{Eq Number}
N(t)=\int \rho(z,t)dz=\sum_p \sum_n |U_{pn}(t)|^2.
\end{equation}
The time-ordered propagator $U_{pn}(t)$ can be numerically calculated by employing the split-operator technique. Therefore, according to Eq.(\ref{Eq Density}), (\ref{Eq Number}) we can compute various properties of the electrons produced under the action of the external potential.

\section{Numerical results}

In this work, we use the Sauter potential well oscillating in both width and depth directions, which is given by
\begin{equation}\label{Eq Well}
\begin{aligned}
V(z,t)&=\frac{V_0(t)}{2}\left\{\text{tanh}\left[\frac{z-D(t)/2}{W}\right]-\text{tanh}\left[\frac{z+D(t)/2}{W}\right]\right\}, \\
V_0(t)&=\frac{V_0}{2}[1-\text{cos}(\omega_0 t+\varphi)], \\
D(t)&=W+\frac{D_0}{2}[1-\text{cos}(\omega_0 t)].
\end{aligned}
\end{equation}
Here $V_0(t)$ and $D(t)$ represents the depth and width of the potential well oscillating periodically over time, respectively.
$V_0$ and $D_0$ are the two oscillation amplitudes, $\omega_0$ is the same oscillation frequency of the width and the depth. The phase difference between them is $\varphi$, which varies over a period of $2\pi$. In this paper, $W=0.3\lambda_C$ is the fixed width of the potential edges, and $\lambda_C=1/c$ is the Compton wavelength.

\begin{figure}[htbp]\suppressfloats
\vskip -9.5cm
\includegraphics[width=15cm]{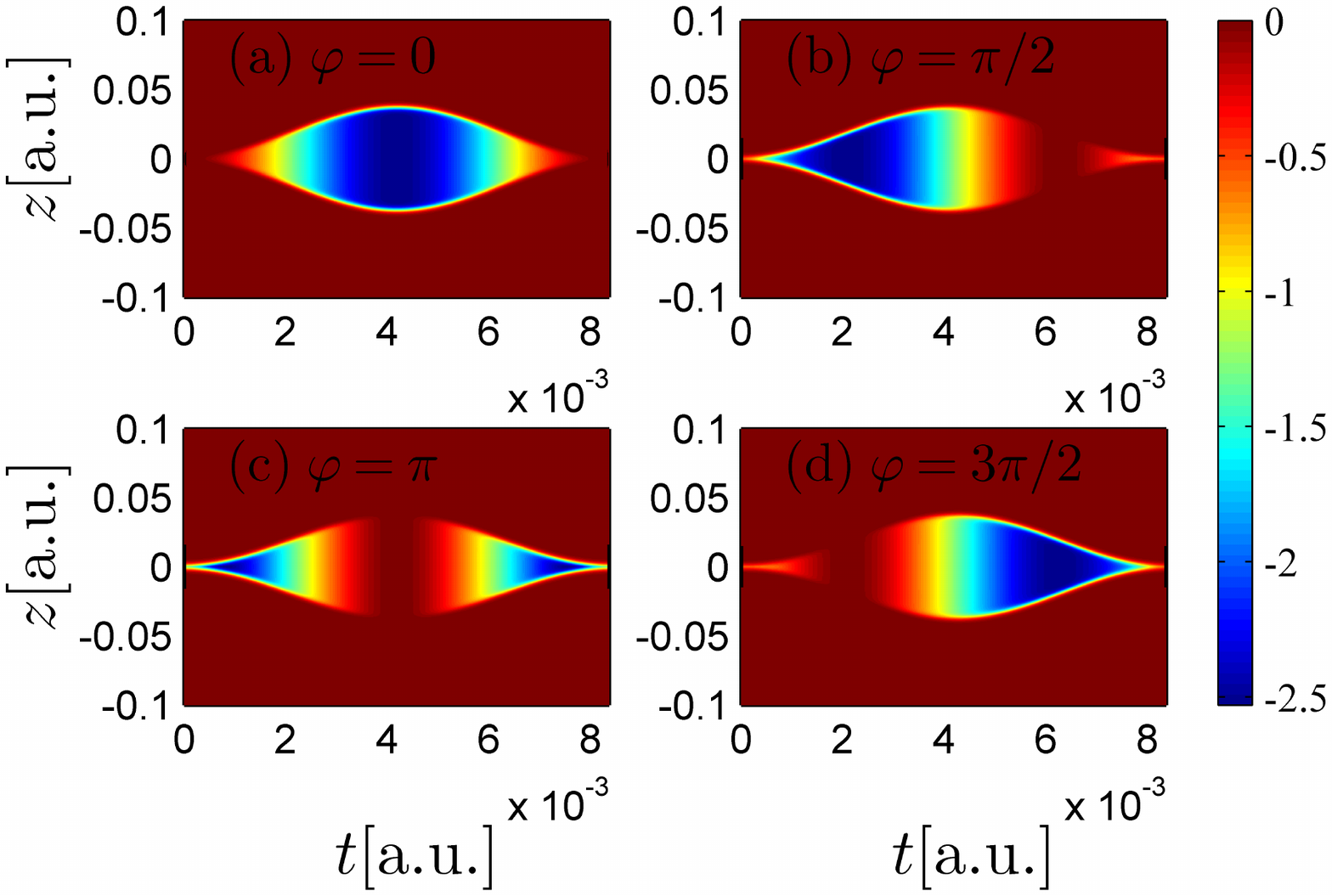}
\caption{\label{fig1} Contour profile plot of the space-time structure of the potential well. (a) is for $\varphi=0$; (b) is for $\varphi=\pi/2$; (c) is for $\varphi=\pi$; (d) is for $\varphi=3\pi/2$. The simulation time is set to $t=50\pi/c^2$. Other parameters are $D_0=10\lambda_C$, $V_0=2.53c^2$, $\omega_0=0.04c^2$. The spatial size is $L=2.5$.}
\end{figure}

From Eq.(\ref{Eq Well}), the time dependence of our model potential well is shown in Fig.\ref{fig1}, where other parameters are set to $D_0=10\lambda_C$, $V_0=2.53c^2$ and $\omega_0=0.04c^2$. The simulation time of the numerical calculation is $t=50\pi/c$, which is less than the time required for the electrons generated in the potential well to leave the simulation regions. The space in the simulation range is from $-1.25$ a.u. to $1.25$ a.u., while we only show the space region from $-0.1$ a.u. to $0.1$ a.u. in the figures. The depth of the potential well is indicated by color. In Fig.\ref{fig1}(a), the width and depth of the potential well increase or decrease synchronously when the phase difference is zero. In other words, when the width of the potential well reaches maximum (minimum), the depth also reaches maximum (minimum). For $\varphi=\pi$, the variation of the width with time is opposite to that of the depth as shown in Fig.\ref{fig1}(c). The two cases of $\varphi=\pi/2$ and $\varphi=3\pi/2$ are shown in Figs.\ref{fig1}(b) and (d), respectively.

\begin{figure}[htbp]\suppressfloats
\vskip -9.5cm
\includegraphics[width=13cm]{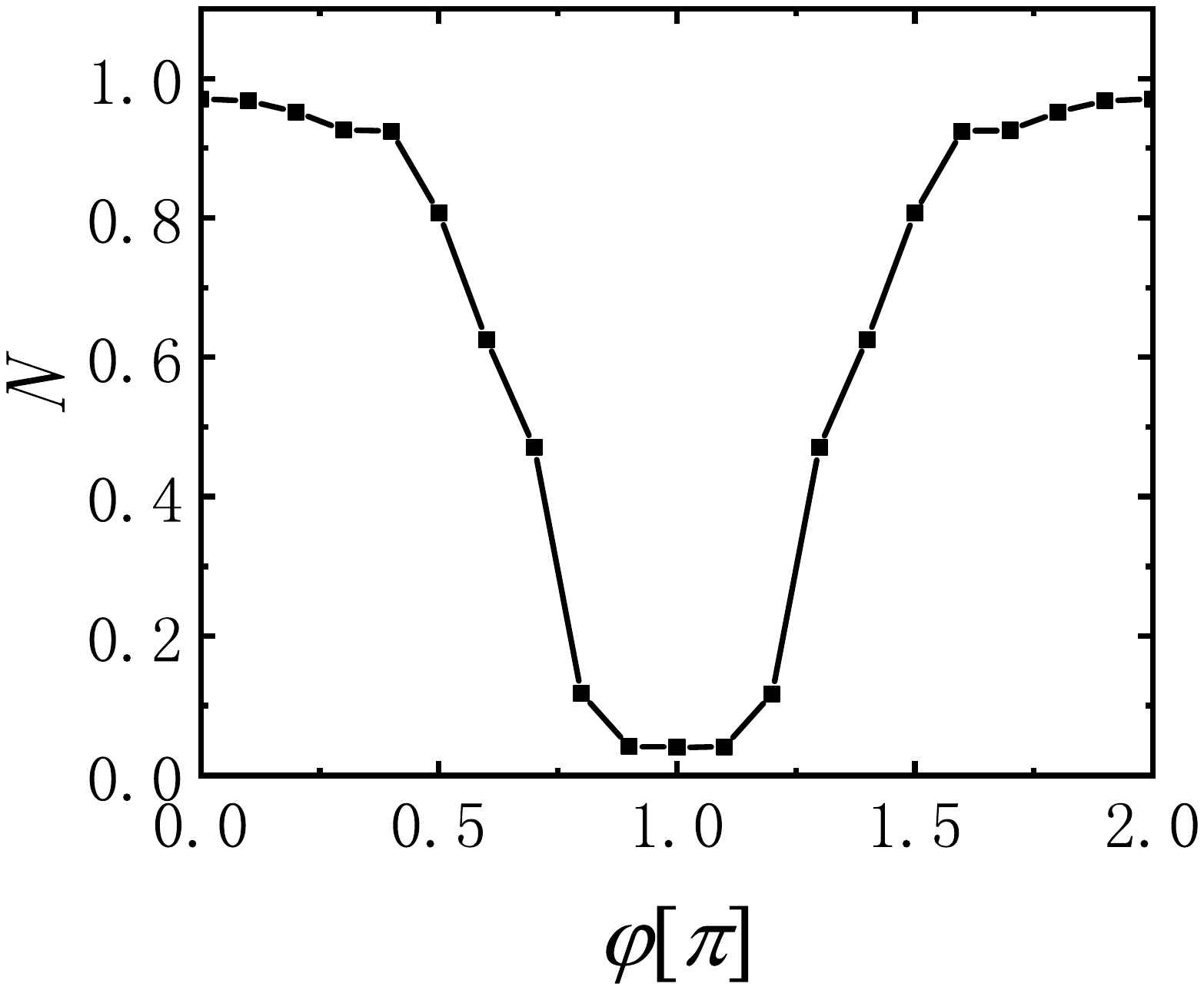}
\caption{\label{fig2} Number of created electrons as a function of phase $\varphi$ over a period of $2\pi$. The simulation time is set to $t=50\pi/c^2$. Other parameters are the same as Fig. \ref{fig1}. }
\end{figure}

 Next we study the effects of the phase difference on the number of created electrons in our model. By solving Eq.(\ref{Eq Number}) numerically, the total number of created electrons and its variation with phase differences over a period of $2\pi$ are represented in Fig.\ref{fig2}. Other parameters are set to $V_0=2.53c^2$, $D_0=10\lambda_C$ and $\omega_0=0.04c^2$. With the increase of phase difference, the number of created electrons decreases firstly and then increases.
 For $\varphi=\pi/2$ and $\varphi=3\pi/2$ corresponding to Fig.\ref{fig1}(b) and (d) respectively, the number of created electrons is the same. As we can see from the figure, the curve is symmetric about $\varphi=\pi$, which can be verified by Eq.(\ref{Eq Well}). Most electrons were produced near $\varphi=0$ or $\varphi=2\pi$, and we define them as the optimal phase differences for $\omega_0=0.04c^2$. The two cases of $\varphi=0$ and $\varphi=2\pi$ all correspond to Fig.\ref{fig1}(a). Intuitively, the wider and deeper the potential well, the more electrons are produced. The minimum number of electrons are created with $\varphi=\pi$, which corresponds to Fig.\ref{fig1}(c). When the width of the potential well is very large(small), the depth is very small(large), which inhibits the generation of electrons.

 To explain above solutions, we need to start with the instantaneous bound state. The following formula can be used to calculate energy levels in our model of potential well \cite{Greiner1985},
\begin{equation}\label{Eq Level}
cp_2(t)\text{cot}(p_2(t)D(t))=E(t)V_0(t)/cp_1(t)-cp_1(t),
\end{equation}
where $p_1(t)=\sqrt{c^2-E(t)^2/c^2}$ and $p_2(t)=\sqrt{(E(t)+V_0(t))^2/c^2-c^2}$. Since the depth and width are all time-dependent functions, other parameters also depend on time.

\begin{figure}[htbp]\suppressfloats
\hskip +15cm
\vskip -20cm
\includegraphics[width=23cm]{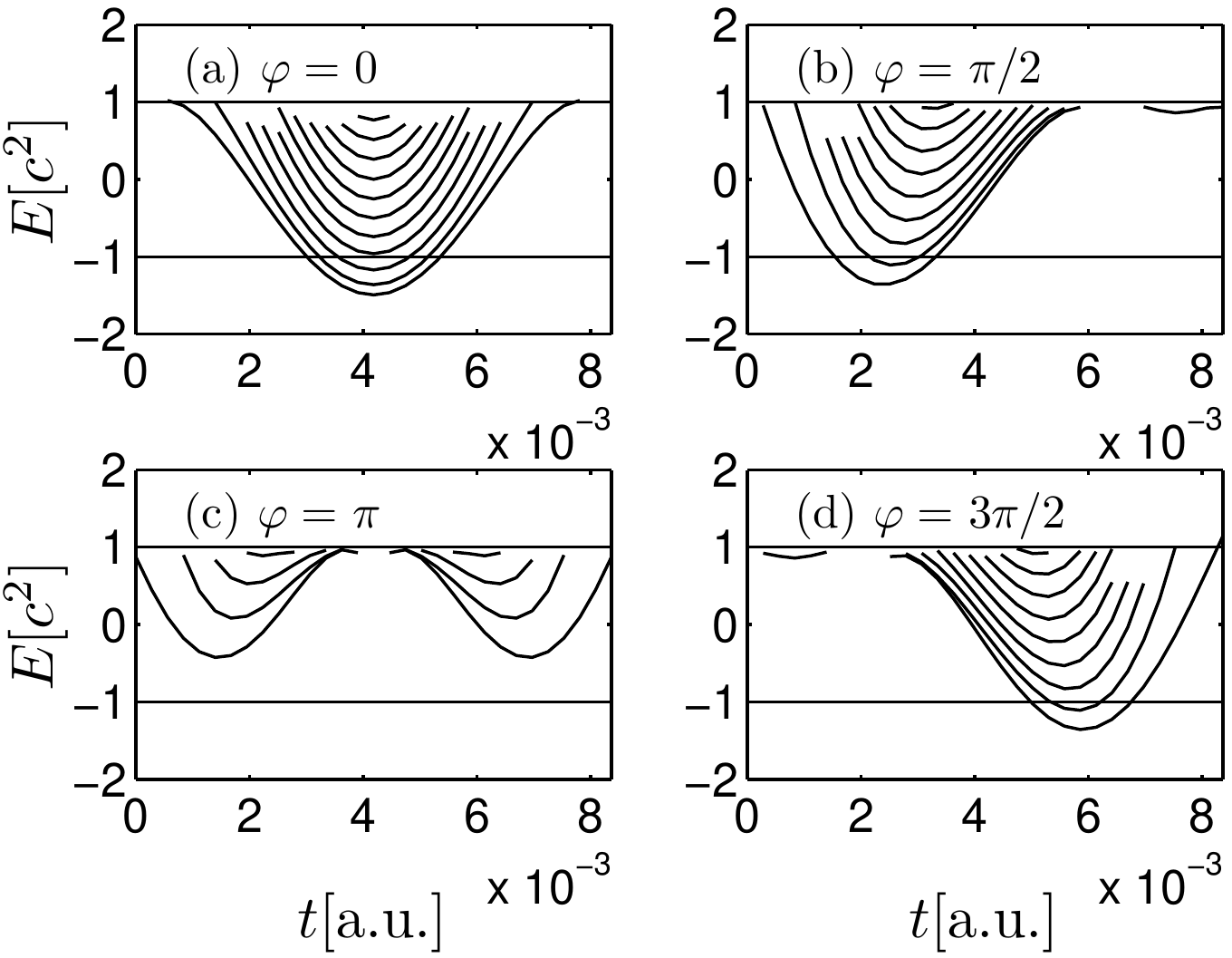}
\caption{\label{fig3} Instantaneous eigenvalues of the potential well over time. Other parameters are the same as those in Fig.\ref{fig1}.}
\end{figure}
Eigenvalues of the potential well over time are presented in Fig.\ref{fig3}. Other parameters are same as those in Fig.\ref{fig1}. By combining with Fig.\ref{fig1}, we can find that the bound states appear at the moment when the potential well is deeper. Furthermore, the deeper and wider the potential well is, the more energy levels there are.
In Fig.\ref{fig3}(a), there are 11 energy levels, the lowest three of which dive into the negative continuum states from $t=3.0\times 10^{-3}$a.u. to $t=5.3\times 10^{-3}$a.u., with a time interval of $\triangle t_1=2.3\times 10^{-3}$a.u..
In Fig.\ref{fig3}(b), the number of energy levels is 9, which is less than that in Fig.\ref{fig3}(a). The lowest two energy levels dive into the negative continuum states during the time interval of $\bigtriangleup t_2=1.7\times 10^{-3}$a.u..
In Fig.\ref{fig3}(c), there are three bound states, none of which dives into the negative continuum states, i.e., all energy levels are trapped in the gap.
In Fig.\ref{fig3}(d), the number of bound states and the dive time interval are all the same as those in Fig.\ref{fig3}(b) except an obvious time-translation mirror symmetry. To facilitate the understanding, we named the time intervals $\triangle t_1=2.3\times 10^{-3}$a.u. and $\bigtriangleup t_2=1.7\times 10^{-3}$a.u. as the efficient interaction time.

 According to the study of Liu \textit{et al.}, the number of created electrons with sufficient interaction time (the critical time) for a potential well that is sufficiently deep and wide is equal to the number of bound states diving into the negative continuum state \cite{Liu2014}. In their paper, for a static Sauter potential well with $D=4.55/c$, $V_0=2.53c^2$ and $W=0.3/c$, the critical time is calculated as $t_{cr}=2.58\times 10^{-3}$a.u. with one bound states diving into the negative continuum. And the critical time is proportional to the number of bound states that are triggered by the static potential.

 Obviously, in one cycle the maximum efficient interaction time $\triangle t_1=2.3\times 10^{-3}$a.u. in Fig.\ref{fig3} is shorter than the critical time $t_{cr}=2.58\times 10^{-3}$a.u.. Therefore, we can infer that the final number of created electrons is less than the number of bound states in the negative continuum state, since the efficient interaction time is not large enough for the electrons to occupy completely. This reasonable inference is confirmed by Fig.\ref{fig2}.
 The final number of created electrons in Fig.\ref{fig2} for $\varphi=0$ is less than three, which is the number of bound states diving into the negative continuum state in Fig.\ref{fig3}(a). For $\varphi=\pi/2$ and $\varphi=3\pi/2$, the same conclusion holds. However, the similar results can be concluded qualitatively, i.e., more numbers of bound state dive into the negative continuum, the more pairs can be created. In short the number of bound states and the effective interaction time depend strongly on the phase difference. It can explain the relation between the number of created pairs and the phase difference. There exists an optimal phase difference where the maximum number can be achieved. Since the effective time depends not only on the phase difference but also on the oscillation frequency, it is also found that the higher the frequency, the shorter the effective interaction time.

\begin{figure}[htbp]\suppressfloats
\vskip -9.5cm
\includegraphics[width=15cm]{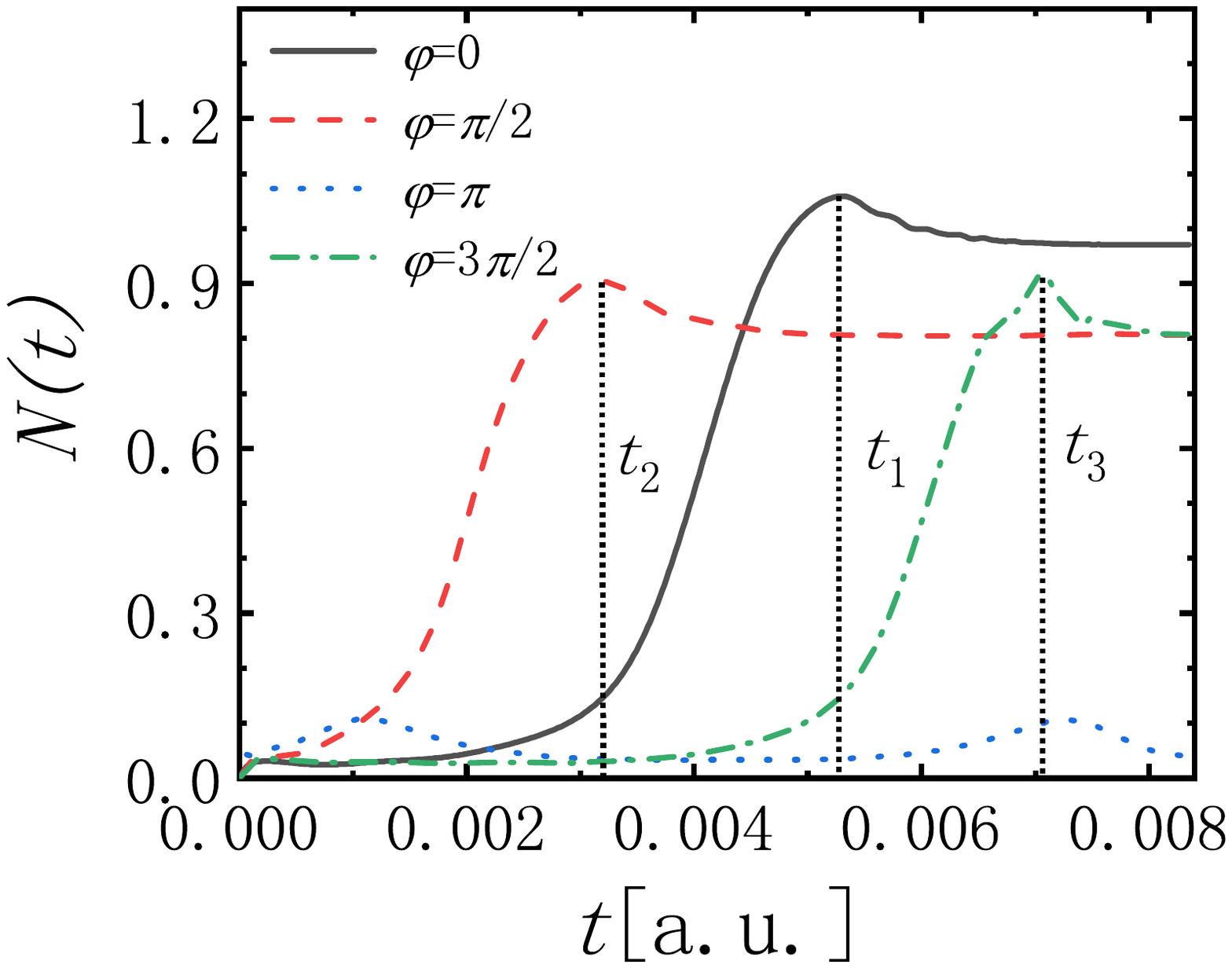}
\caption{\label{fig4} Number of created electrons from $t=0$ to $t=50\pi/c^2$ for phase difference $\varphi=0$ (the black solid line), $\varphi=\pi/2$ (red double dash line), $\varphi=\pi$ (blue dotted line) and $\varphi=3\pi/2$ (green dot and dash line). Other parameters are same as those in Fig. \ref{fig1}. }
\end{figure}

To better understand the studied problem, the time evolution of the number of created electrons is shown in Fig.\ref{fig4}.
For $\varphi=0$, the number stops increasing at $t_1=5.3\times 10^{-3}$a.u., and stabilizes at $N=0.97$ after a brief decrease. The decrease is due to the annihilation between electron and positron. Compared with Fig.\ref{fig3}(a), it is found that the time when the number of created electrons stops increasing corresponds to that time when the lowest bound state leaves the negative continuum state. Moreover, the time region with higher growth rate in this figure corresponds to the efficient interaction time in Fig.\ref{fig3}(a). This result is also suitable for cases of other phase differences.

For $\varphi=\pi/2$ in Fig.\ref{fig4}, the maximum value of the red double dash curve is approximately at $t_2=3.3\times 10^{-3}$a.u., which is the same as the time for the lowest bound state to leave the negative continuum state in Fig.\ref{fig3}(b).
For $\varphi=\pi$, the positions of the two peaks in the blue dotted line of Fig.\ref{fig4} correspond to those positions of the two valleys in Fig.\ref{fig3}(c).

For $\varphi=3\pi/2$, the peak is located at $t_3=7.1\times 10^{-3}$a.u. corresponding approximately to the time for the lowest bound state to leave the negative continuum state in Fig.\ref{fig3}(d). This reflects that the bound states in the negative continuum play the main role in the progress of positron-electron pairs, which has been verified by Liu \textit{et al.} \cite{Liu2014}.
Although these lines in Fig.\ref{fig4} increase at different starting times, these final numbers are consistent with those in Fig.\ref{fig2}.
For $\varphi=\pi/2$ and $\varphi=3\pi/2$, the final numbers of created electrons all tend to $N=0.8$ at the end of the period, which is lower than $N=0.97$ for the case of $\varphi=0$.

So far, we come to the conclusion that the number of electrons is mainly determined by the bound state and the effective interaction time for low frequency oscillation, and the phase difference changes the bound state and the effective interaction time by controlling the depth and width of the potential well. When the oscillation frequency is set to $\omega_0=0.04c^2$, the optimal phase difference is $\varphi=0$.

\begin{figure}[htbp]\suppressfloats
\vskip -9.5cm
\includegraphics[width=15cm]{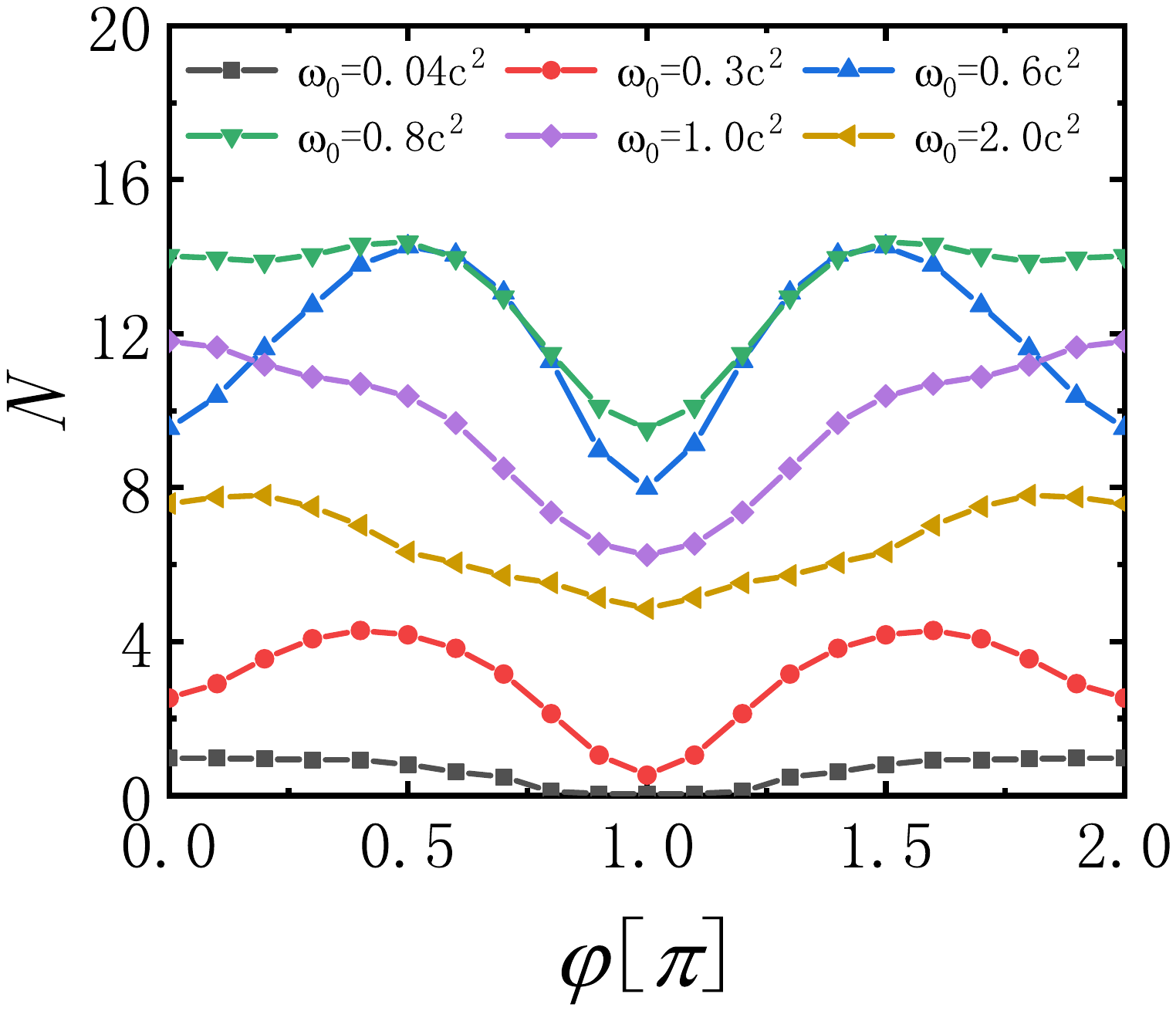}
\caption{\label{fig5} The number of created electrons and its variation with phase $\varphi$ over a period $2\pi$ for different oscillation frequencies.}
\end{figure}

The number of created electrons and its variation with phase $\varphi$ over a period of $2\pi$ for different oscillation frequencies is shown in Fig.\ref{fig5}. As the frequency increases, the curve moves up to a certain height and then comes down. That is to say, with the increase of oscillation frequency, the number of produced electrons does not always increase, which conforms to the conclusion of Ref.\cite{Jiang2013}. Interestingly, the optimal phase difference varies with the increase of frequency. The optimal phase difference is located at $\varphi=0$, $0.4\pi$, $0.5\pi$, $0.5\pi$, $0$ and $0.2\pi$ for $\omega_0=0.04c^2$, $0.3c^2$, $0.6c^2$, $0.8c^2$, $1.0c^2$ and $2.0c^2$, respectively.
Note that, the figure is symmetric with respect to $\varphi=\pi$, which is the phase difference corresponding to the minimum number of created electrons.
When the frequency is set to $\omega_0=0.8c^2$, the most electrons are created on conditions that $0<\varphi<\pi/2$ or $3\pi/2<\varphi<2\pi$. The maximum number is about $14$, which can also be obtained by lowering the oscillation frequency to $\omega_0=0.6c^2$ and changing the phase difference to $\varphi=0.5\pi$.

\begin{figure}[htbp]\suppressfloats
\vskip -9.5cm
\includegraphics[width=15cm]{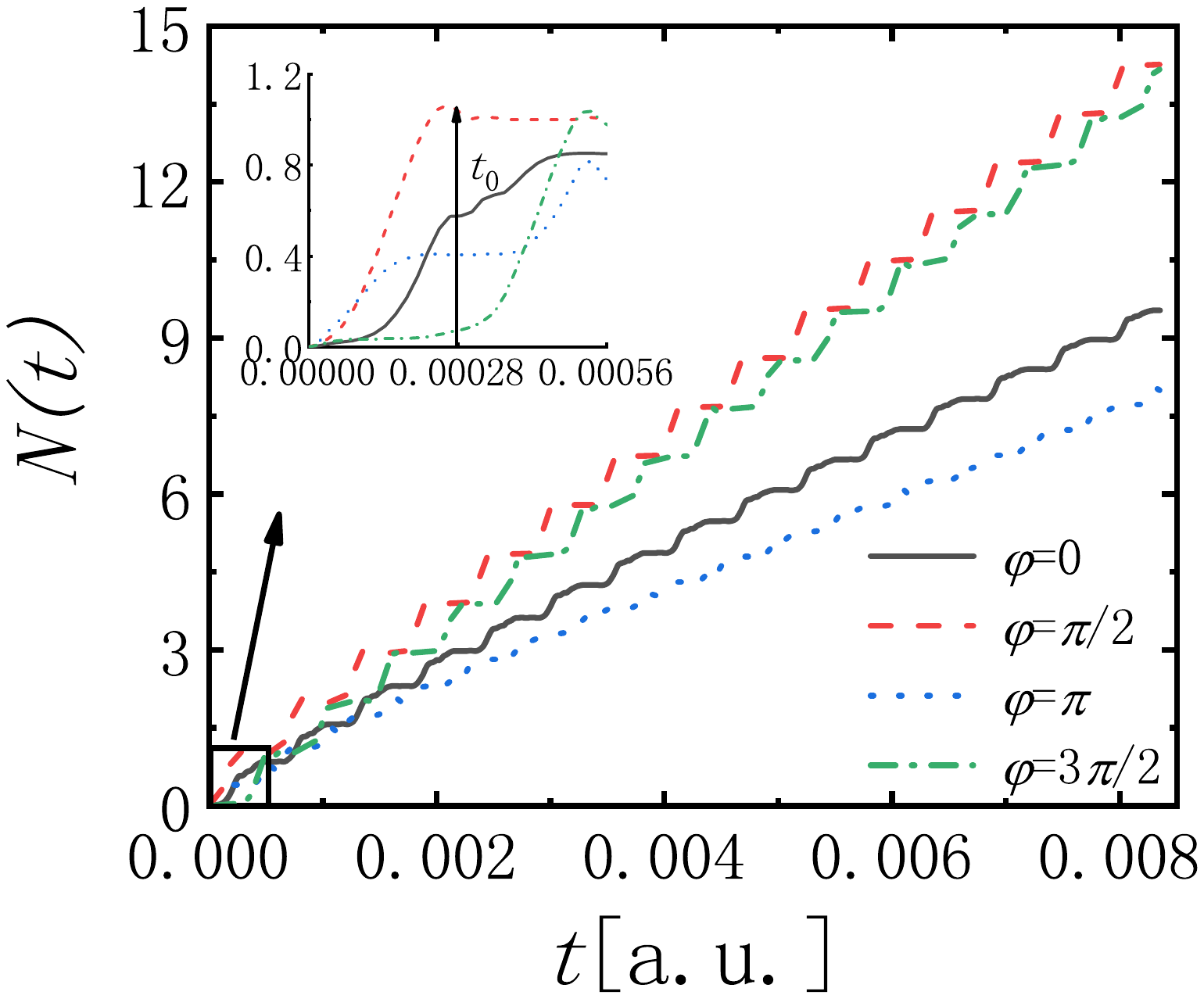}
\caption{\label{fig6} Time evolution of the number of created electrons from $t=0$ to $t=50\pi/c^2$ for $\varphi=0$ (the black solid line), $\varphi=\pi/2$ (red double dash line), $\varphi=\pi$ (blue dotted line) and $\varphi=3\pi/2$ (green dot and dash line). Other parameters are the same as those in Fig. \ref{fig1}, except the frequency $\omega_0=0.6c^2$(15 cycles).}
\end{figure}

The time evolution of the created number with $\omega_0=0.6c^2$ is shown in Fig.\ref{fig6}. The period of oscillation is set to $T=2\pi/\omega_0\approx 5.58\times 10^{-4}$a.u.. Because of the robustness characteristic, the number of created electrons increases stepwise with the simulation time during fifteen cycles \cite{Liu2014}.
Unlike in Fig.\ref{fig4}, the final number is no longer the largest for $\varphi=0$. For the two cases of $\varphi=\pi/2$ and $\varphi=3\pi/2$, the two curves are periodically separated and intersected with the final number being large than those for $\varphi=\pi/2$ and $\varphi=3\pi/2$. Since every curve repeats the same pattern of variation rule in each period, we choose the first oscillation period to amplify and analyze it, as shown in the inset.

Comparing with Fig.\ref{fig4}, the effective interaction time in Fig.\ref{fig6} are shortened to $\triangle t_1/15=1.5\times 10^{-4}$a.u. for $\varphi=0$,  $\triangle t_2/15=1.1\times 10^{-4}$a.u. for $\varphi=\pi/2$ and $\varphi=3\pi/2$. According to the previous analysis, the decrease of the effective interaction time will reduce the number of electrons. But the final number on the case of $\varphi=\pi$ in the end of the first period is $0.82$ in Fig.\ref{fig6}, which is much larger than that in Fig.\ref{fig4}. The method of the transient bound state can not fully explain this result. There may be some other effects such as the multi-photon process and the Pauli blocking that have more significant impacts on the creation of electrons. By absorbing multiple laser photons, electrons with negative energy in the negative continuum state can transit to the positive continuum to create electron-positron pairs. Moreover, the bound states in the gap provide some ladders for the electron transition, which enhances the probability of the electron transition. For the case of $\varphi=0$ in the inset, the number of created electrons stops increasing when $t_0=2.8\times 10^{-4}$a.u.. In the second half of the cycle, the number of electrons is suppressed, which may be caused by the Pauli blocking.

When the oscillation frequency is very high, the electrons generated on both sides of the potential well move very fast. Supposing they move toward the middle with the speed of light, the encounter time would be at $t_0=D/2c$. The maximum time for electrons created in the middle of a cycle to meet is about $2.7\times 10^{-4}$a.u., which is equal to half of the period. According to the Pauli exclusion principle, each bound state can only be occupied by one electron. Thus we can guess that electrons produced in the first half of the period are trapped in the potential well, occupying these bound states, so that the electrons generated in the second half of the period have no bound states available. For the cases of $\varphi=\pi/2$ and $\varphi=3\pi/2$, the final number in the inset is $1.0$, which is larger than $0.8$ in Fig.\ref{fig4}. This is because that the Pauli blocking is perfectly avoided for the two cases, and the multi-photon process increases the probability of creation compared with that in Fig.\ref{fig4}.

\begin{figure}[htbp]\suppressfloats
\vskip -9.5cm
\includegraphics[width=15cm]{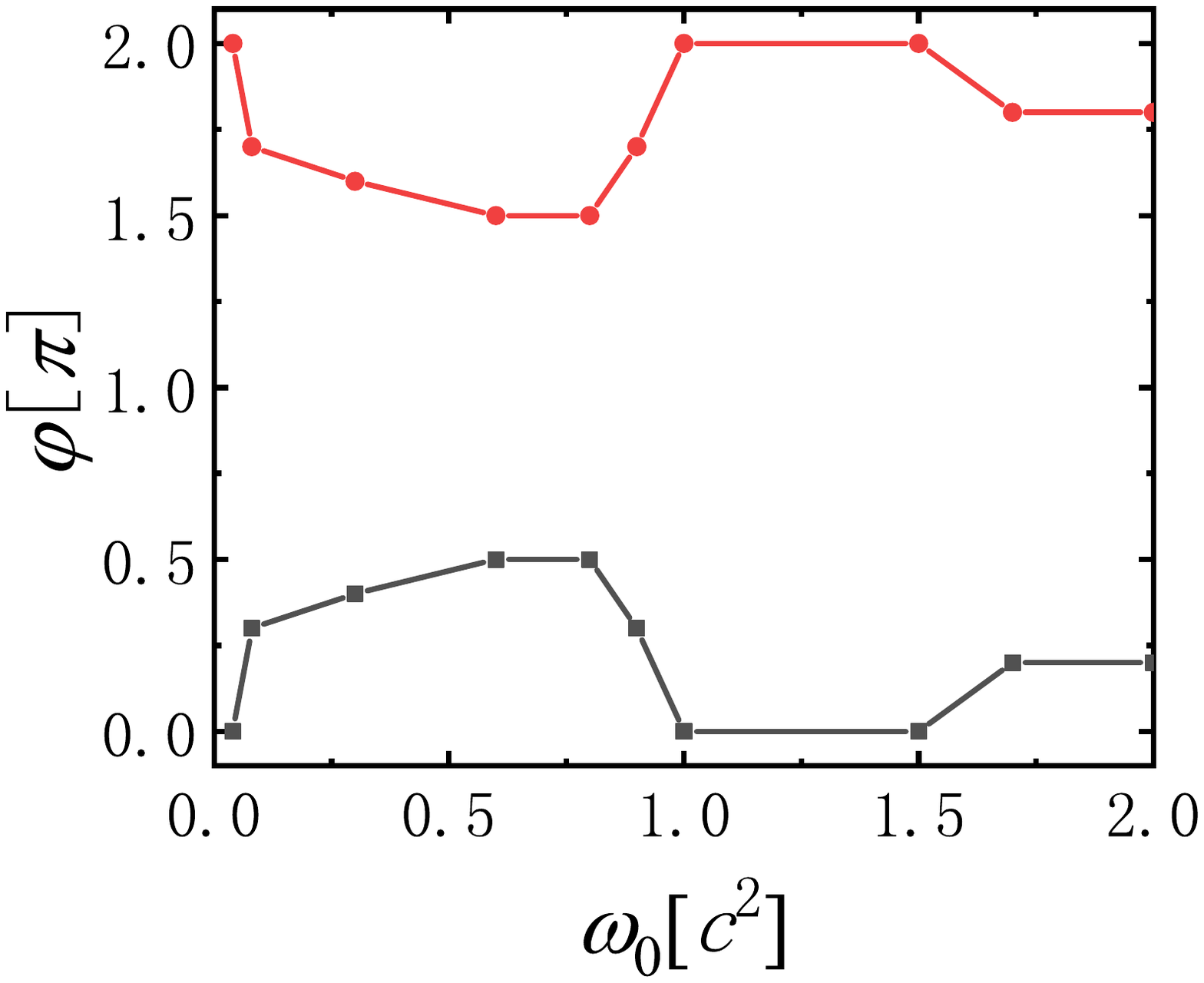}
\caption{\label{fig7} The optimal phase difference and its variation with oscillation frequency in a period of $2\pi$.}
\end{figure}
The optimal phase difference in a period of $2\pi$ and its variation with oscillation frequency is presented in Fig.\ref{fig7}. Since the two lines are symmetric about $\varphi=\pi$, we only analyze the lower one. The optimal phase difference changes non-monotonically with the increase of oscillation frequency.
For the oscillation frequency ranging from $0.04c^2$ to $1.0c^2$, the optimal phase difference increases slowly to $\varphi=\pi/2$ and then decreases to $0$. When the oscillation frequency ranging from $1.0c^2$ to $2.0c^2$, the optimal phase difference located near $\varphi=0$.

\begin{table}[htbp]	\label{Table1}  
	\centering  
	\caption{The maximum, the minimum number of created electrons and the ratio between them for different frequencies in Fig.\ref{fig5}}  
    \renewcommand\arraystretch{1.0}  
    \renewcommand\tabcolsep{10.0pt} 
    \begin{tabular}{l c c c}
          \hline \hline 
          $\omega_0(c^2)$ & $N_{max}$ & $N_{min}$ & $R(N_{max}/N_{min})$\\  
          \hline 
          0.04&0.97($\varphi=0$)&0.04($\varphi=\pi$)&24.25 \\
          0.3&4.29($\varphi=\pi/2$)&0.53($\varphi=\pi$)&8.09 \\
          0.6&14.26($\varphi=\pi/2$)&7.99($\varphi=\pi$)&1.78 \\
          0.8&14.38($\varphi=\pi/2$)&9.52($\varphi=\pi$)&1.51 \\
          1.0&11.81($\varphi=0$)&6.24($\varphi=\pi$)&1.89 \\
          2.0&7.80($\varphi=\pi/5$)&4.85($\varphi=\pi$)&1.61 \\
          \hline \hline
    \end{tabular}
\end{table}
Table I shows the maximum, the minimum number of created electrons and the ratio between them for different frequencies in Fig.\ref{fig5}. From the second column, the optimal phase difference ranges from 0 to $\pi/2$. With the increase of frequency, the parameter $N_{max}$ increases to maximum 14.38 with $\omega_0=0.8c^2$ and then decreases.
In the third column, the parameter $N_{min}$ also increases to maximum 9.52 with $\omega_0=0.8c^2$ and then decreases. The corresponding phase difference is constant $\pi$. The ratio of the maximum to the minimum is larger for a lower frequency, and it is less than 2 for $\omega_0>0.6c^2$.
 What we can take as an important reference from the table is that for the three cases that the parameter $N_{max}$ is greater than 10, the corresponding ranges of the frequency and the phase difference are $0.6c^2<\omega_0<1.0c^2$ and $0<\varphi<\pi/2$ respectively.

\section{Conclusion}

We have investigated the effects of the phase difference between the depth and width oscillation on electron-positron pair creation employing the computational quantum field theory. Firstly, we select a slowly changing field and simulated the number of created electrons and its variation with phase difference. Secondly, we studied the phase modulation for different oscillation frequencies. The main results are as follows.

1. For low-frequency oscillations, the phase difference has a great effect on the number of created pairs. When the width and depth of potential well changes simultaneously($\varphi=0$), the number of created pairs reaches maximum. Conversely, when the phase difference is set to $\varphi=\pi$, there are few pairs created.

2. With the increase of the frequency, the variation of the optimal phase parameter is nonmonotonic.
On the condition that $\omega_0<1.0c^2$, the optimal phase parameter is mainly located at $\varphi=\pi/2$. For $1.0c^2<\omega_0<2.0c^2$, it is mostly around $\varphi=0$.

3. By controlling the frequency $\omega_0$ ranging from $0.6c^2$ to $1.0c^2$ and the phase difference $\varphi$ from 0 to $\pi/2$, more electron-positron pairs can be obtained.

When the potential well changes slowly, the results can be explained by the instantaneous bound states. For the higher frequency case, the multi-photon effect is enhanced. Because of the robustness character, the number of created pairs rises stepwise over time. For the case of simultaneous variation of the width and depth of potential well, the Pauli blocking effect has a strong inhibitory effect on creation of electrons at high frequencies.

In this work, we study with the same frequency parameter in the width and the depth oscillations. If the frequencies of the two oscillations are different, there may be more abundant pairs production. Some of other factors deserve to be further considered and studied in future, which is beyond the scope of present paper.
Above conclusions may have possible reference value to laboratory experiments in the generation of electron-positron pairs.

\begin{acknowledgments}
We thank Dr. Q Su for helpful discussions. This work was supported by the National Natural Science Foundation of China (NSFC) under Grants No. 11875007 and No. 11475026. The computation was carried out at the
HSCC of the Beijing Normal University.
\end{acknowledgments}


\begin{thebibliography}{99}\suppressfloats
\bibitem{Dirac1928}
P. A. M. Dirac, Proc. Roy. Soc. Lond. A \textbf{117}, 610(1928).

\bibitem{Anderson1933}
C. D. Anderson, Phys. Rev. \textbf{43}, 491(1933).

\bibitem{Sauter1931}
F. Sauter, Z. Phys. \textbf{69}, 742(1931).

\bibitem{Heisenberg1936}
W. Heisenberg and H. Euler, Z. Phys. \textbf{98}, 714(1936).

\bibitem{Schwinger1951}
J. Schwinger, Phys. Rev. \textbf{82}, 664(1951).

\bibitem{Brezin1970}
E. Brezin and C. Itzykson, Phys. Rev. D \textbf{2}, 1191(1970).

\bibitem{Marinov1977}
M. S. Marinov and V. S. Popov, Fort. Phys. \textbf{25}, 373(1977).

\bibitem{Gies2005}
H. Gies and K. Klingm$\rm{\ddot{u}}$ller, Phys. Rev. D \textbf{72}, 065001(2005).

\bibitem{Dunne2006}
G. V. Dunne, Q. H. Wang, H. Gies, and C. Schubert, Phys. Rev. D \textbf{73}, 065028(2006).

\bibitem{Ilderton2015}
A. Ilderton, G. Torgrimsson, and  J. W{\aa}rdh, Phys. Rev. D \textbf{92}, 065001(2015).

\bibitem{Schneider2016}
C. Schneider and R. Sch$\rm{\ddot{u}}$tzhold, JHEP \textbf{2}, 164(2016).

\bibitem{Li2015}
Z. L. Li, D. Lu, and B. S. Xie, Phys. Rev. D \textbf{92}, 085001(2015).

\bibitem{Blinne2016}
A. Blinne and E. Strobel, Phys. Rev. D \textbf{93}, 025014(2016).

\bibitem{Kohlfurst2018}
C. Kohlf$\rm{\ddot{u}}$rst and R. Alkofer, Phys. Rev. D \textbf{97}, 036026(2018).

\bibitem{Olugh2019}
O. Olugh, Z. L. Li, B. S. Xie, and R. Alkofer, Phys. Rev. D \textbf{99}, 036003(2019).

\bibitem{Kluger1991}
Y. Kluger, J. M. Eisenberg, B. Svetitsky, F. Cooper, and E. Mottola, Phys. Rev. Lett. \textbf{67}, 2427(1991).

\bibitem{Alkofer2001}
R. Alkofer, M. B. Hecht, C. D. Roberts, S. M. Schmidt, and D. V. Vinnik, Phys. Rev. Lett. \textbf{87}, 193902(2001).

\bibitem{JiangMin2013}
M. Jiang, B. S. Xie, H. B. Sang, and Z. L. Li, Chin. Phys. B \textbf{22}, 100307(2013).

\bibitem{Sitiwaldi2017}
I. Sitiwaldi and B. S. Xie, Phys. Lett. B \textbf{768}, 174(2017).

\bibitem{Krekora2005}
P. Krekora, K. Cooley, Q. Su, and R. Grobe, Phys. Rev. Lett. \textbf{95}, 070403(2005).

\bibitem{Lv2013}
Q. Z. Lv, Y. Liu, Y. J. Li, R. Grobe, and Q. Su, Phys. Rev. Lett. \textbf{111}, 183204(2013).

\bibitem{Tang2013}
S. Tang, B. S. Xie, D. Lu, H. Y. Wang, L. B. Fu, and J. Liu, Phys. Rev. A \textbf{88}, 012106(2013).

\bibitem{Jiang2013}
M. Jiang, Q. Z. Lv, Z. M. Sheng, R. Grobe, and Q. Su, Phys. Rev. A \textbf{87}, 042503(2013).

\bibitem{Jiang2014}
M. Jiang, Q. Z. Lv, Y. Liu, R. Grobe, and Q. Su, Phys. Rev. A \textbf{90}, 032101(2014).

\bibitem{Liu2014}
Y. Liu, M. Jiang, Q. Z. Lv, Y. T. Li, R. Grobe, and Q. Su, Phys. Rev. A \textbf{89}, 012127(2014).

\bibitem{Gong2018}
C. Gong, Z. L. Li, and Y. J. Li, Phys. Rev. A \textbf{98}, 043424(2018).

\bibitem{Abdukerim2013}
N. Abdukerim, Z. L. Li, and B. S. Xie, Phys. Lett. B \textbf{726}, 820(2013).

\bibitem{Kohlfurst2016}
C. Kohlf$\rm{\ddot{u}}$rst and R. Alkofer, Phys. Lett. B \textbf{756}, 371(2016).

\bibitem{Liu2015}
Y. Liu, Q. Z. Lv, Y. T. Li, R. Grobe, and Q. Su, Phys. Rev. A \textbf{91}, 052123(2015).

\bibitem{Greiner1985}
W. Greiner, B. M$\rm{\ddot{u}}$ller, and J. Rafelski, \textsl{Quantum Electrodynamics of Strong Fields} (Springer Verlag, Berlin, 1985).

\end{thebibliography}
\end{document}